\documentclass[sigconf]{acmart}
\AtBeginDocument{%
  \providecommand\BibTeX{{%
    \normalfont B\kern-0.5em{\scshape i\kern-0.25em b}\kern-0.8em\TeX}}}

\setcopyright{acmlicensed}
\copyrightyear{2025}
\acmYear{2025}
\setcopyright{acmlicensed}\acmConference[KDD '25]{Proceedings of the 31st ACM SIGKDD Conference on Knowledge Discovery and Data Mining V.2}{August 3--7, 2025}{Toronto, ON, Canada}
\acmBooktitle{Proceedings of the 31st ACM SIGKDD Conference on Knowledge Discovery and Data Mining V.2 (KDD '25), August 3--7, 2025, Toronto, ON, Canada}
\acmDOI{10.1145/3711896.3737202}
\acmISBN{979-8-4007-1454-2/2025/08}

\usepackage{textcomp}
\usepackage{colortbl}
\usepackage{float}
\usepackage{graphicx}
\usepackage{caption}
\usepackage{afterpage}
\usepackage{listings}
\usepackage{xcolor}
\usepackage{algorithm}
\usepackage{algpseudocode}
\usepackage{hyperref}
\usepackage{xspace}
\usepackage{booktabs}
\usepackage{tabularx}
\usepackage{enumitem}
\usepackage{xurl}
\usepackage{pifont}

\newcommand{\method}{\textit{ChatPD}\xspace}

\colorlet{punct}{red!60!black}
\definecolor{background}{HTML}{EEEEEE}
\definecolor{delim}{RGB}{20,105,176}
\colorlet{numb}{magenta!60!black}
\definecolor{Gray1}{gray}{0.90}

\lstdefinelanguage{json}{
    basicstyle=\footnotesize\ttfamily,
    numberstyle=\scriptsize,
    stepnumber=1,
    numbersep=8pt,
    showstringspaces=false,
    breaklines=true,
    frame=lines,
    backgroundcolor=\color{background},
    literate=
     *{0}{{{\color{numb}0}}}{1}
      {1}{{{\color{numb}1}}}{1}
      {2}{{{\color{numb}2}}}{1}
      {3}{{{\color{numb}3}}}{1}
      {4}{{{\color{numb}4}}}{1}
      {5}{{{\color{numb}5}}}{1}
      {6}{{{\color{numb}6}}}{1}
      {7}{{{\color{numb}7}}}{1}
      {8}{{{\color{numb}8}}}{1}
      {9}{{{\color{numb}9}}}{1}
      {:}{{{\color{punct}{:}}}}{1}
      {,}{{{\color{punct}{,}}}}{1}
      {\{}{{{\color{delim}{\{}}}}{1}
      {\}}{{{\color{delim}{\}}}}}{1}
      {[}{{{\color{delim}{[}}}}{1}
      {]}{{{\color{delim}{]}}}}{1},
}

\usepackage[skip=2pt]{caption}
\def\BibTeX{{\rm B\kern-.05em{\sc i\kern-.025em b}\kern-.08em
    T\kern-.1667em\lower.7ex\hbox{E}\kern-.125emX}}

\begin{document}

\title{ChatPD: An LLM-driven Paper-Dataset Networking System}

\author{Anjie Xu}
\affiliation{%
  \institution{Key Lab of High Confidence Software Technologies (Peking University), Ministry of Education}
  \city{Beijing}
  \country{China}
}
\affiliation{%
  \institution{School of Computer Science, Peking University}
  \city{Beijing}
  \country{China}
}
\email{anjiexu@stu.pku.edu.cn}

\author{Ruiqing Ding}
\affiliation{%
  \institution{Key Laboratory of Process Optimization and Intelligent Decision-making, \\Ministry of Education}
  \city{Anhui}
  \country{China}
}
\affiliation{%
  \institution{School of Management, Hefei University of Technology}
  \city{Anhui}
  \country{China}
}
\email{ruiqingding@hfut.edu.cn}

\author{Leye Wang}
\authornote{Corresponding author.}
\affiliation{%
  \institution{Key Lab of High Confidence Software Technologies (Peking University), Ministry of Education}
  \city{Beijing}
  \country{China}
}
\affiliation{%
  \institution{School of Computer Science, Peking University}
  \city{Beijing}
  \country{China}
}
\email{leyewang@pku.edu.cn}

\begin{abstract}

Scientific research heavily depends on suitable datasets for method validation, but existing academic platforms with dataset management like PapersWithCode suffer from inefficiencies in their manual workflow. To overcome this bottleneck, we present a system, called ChatPD, that utilizes Large Language Models (LLMs) to automate dataset information extraction from academic papers and construct a structured paper-dataset network. Our system consists of three key modules: \textit{paper collection}, \textit{dataset information extraction}, and \textit{dataset entity resolution} to construct paper-dataset networks. Specifically, we propose a \textit{Graph Completion and Inference} strategy to map dataset descriptions to their corresponding entities. Through extensive experiments, we demonstrate that ChatPD not only outperforms the existing platform PapersWithCode in dataset usage extraction but also achieves about 90\% precision and recall in entity resolution tasks. Moreover, we have deployed ChatPD to continuously extract which datasets are used in papers, and provide a dataset discovery service, such as task-specific dataset queries and similar dataset recommendations. We open source ChatPD and the current paper-dataset network on this \href{https://github.com/ChatPD-web/ChatPD}{GitHub repository}.

\end{abstract}




\maketitle

\section{Introduction}

In scientific research, datasets play a crucial role in method validation, model training, and result evaluation. Currently, research in many fields relies heavily on datasets, such as disease prediction in the medical field \cite{goldstein2017opportunities} and climate forecasting meteorology \cite{olaiya2012application}.
These studies often involve diverse datasets that span multiple disciplines. 
While researchers may be familiar with datasets within their own field, they may not be aware of datasets from other disciplines that could be beneficial to their research, which makes the process of finding a suitable dataset challenging.

Despite the maturity of information retrieval technologies in the text domain, numerous challenges persist in the realm of data retrieval \cite{gregory2020lost,koesten2017trials}. These challenges stem from the inherent complexity of datasets, which often come in diverse formats such as images, videos, and structured tables, making traditional text-based retrieval methods insufficient. Moreover, the lack of comprehensive and standardized metadata provided by data publishers creates further barriers to locating relevant datasets. As a result, users frequently struggle to find suitable datasets for their research~\cite{chapman2020dataset}.

To make dataset discovery more efficient, we aim to profile the usage of datasets in academic papers and construct a structured paper-dataset network. This network can provide a better understanding of dataset impact, foster reproducibility, and improve dataset discoverability for future research. While some academic platforms, like Google Dataset Search~\cite{brickley2019google} and PapersWithCode (PwC)~\cite{martinez2021research}, have made progress in linking papers to datasets, they still rely heavily on manual annotation or rule-based methods, which are time-consuming and error-prone. Furthermore, existing methods frequently fail to capture fine-grained dataset attributes critical to researchers – including data types, size, and specific usage contexts – limiting their utility in comprehensive research analysis.

In this paper, we introduce \textbf{\method}, a novel system that leverages Large Language Models (LLMs) to automate the construction of a \textit{\textbf{P}}aper-\textit{\textbf{D}}ataset network. We design a dataset information template based on aspects that researchers usually focus on when studying datasets~\cite{koesten2020everything}, and incorporate LLMs to analyze academic papers and extract dataset-related information. While LLMs generate large amounts of textual output at low cost, to integrate this data with existing academic platforms like PwC, we develop an algorithm based on graph completion and inference to map textual descriptions of datasets to the corresponding dataset entities in the dataset database, tailored to the characteristics of our data. Through our system, we obtain a high-quality paper-dataset network with rich metadata information about datasets, which can be used for dataset discovery and recommendation. Finally, we deploy \method as a practical dataset discovery service on \textbf{\url{https://chatpd-web.github.io/chatpd-web}}, supporting regular construction of AI-related paper-datasets networks on arXiv.

In summary, our work has the following contributions:

1. We propose \method, an LLM-driven system designed to automatically construct a paper-dataset network. The system is deployed as an online service that supports dataset-related queries, recommendations, and additional functionalities.

2. We comprehensively evaluate the reliability of \method from the perspective of dataset information extraction and entity resolution. For dataset information extraction, \method achieves a precision of $\sim$0.99, significantly surpassing PwC’s result of $\sim$0.83. In entity resolution, \method attains an F1 score of $\sim$0.88, outperforming state-of-the-art entity resolution algorithms \cite{lacoste2013sigma,wu2020zeroer}, which achieve only $\sim$0.68.  

3. By collecting papers on arXiv cs.AI from 2018 to 2024, we have built a continuously evolving paper-dataset network, which currently includes 60,126 papers, 4,224 dataset entities, and 137,004 paper-dataset usage records.
Notably, the network constructed by \method includes 444 new datasets not covered in PwC, demonstrating the superiority of its automated dataset collection strategy over the manual annotation-based approach employed by PwC.
We open source \method and the collected paper-dataset network on GitHub: \textbf{\url{https://github.com/ChatPD-web/ChatPD}}.

\begin{figure*}[t]
	\centering
	\includegraphics[width=.9\textwidth]{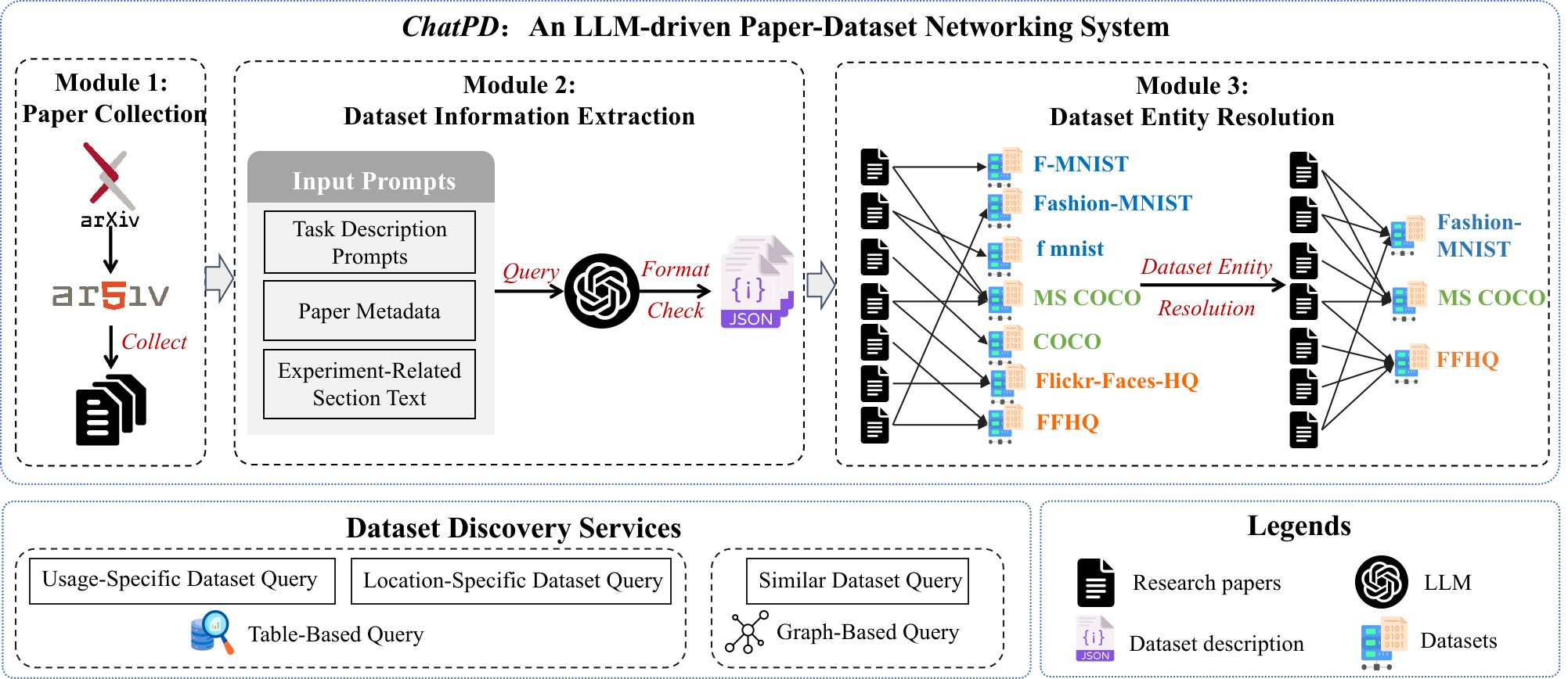}
	\caption{System Architecture of \method.}
	\label{fig:system_arch}
\end{figure*}

\section{Background and Related Work}

Constructing a network that connects papers and datasets to facilitate dataset discovery poses two primary challenges. Firstly, we need to extract pertinent information from scholarly articles. Secondly, given that different papers may refer to the same dataset using diverse names, we are required to perform entity resolution. This process involves mapping varying dataset descriptions to their appropriate entities, enhancing the network's quality.

\subsection{Dataset Discovery}

Dataset discovery is the process of locating, examining, and accessing relevant and valuable datasets for analysis, research, or other purposes.
The retrieval systems for datasets usually rely on the context provided by dataset publishers~\cite{chapman2020dataset}. Kern et al.~\cite{kern2015there} point out the pivotal role of metadata in the discovery of datasets. 
Following this idea, various studies have contributed to the development of dataset summaries and metadata to enhance dataset retrieval~\cite{grusky2018newsroom,wang2022textomics, koesten2020everything}.

Various platforms have been developed to facilitate dataset discovery. Google Dataset Search~\cite{brickley2019google} employs an automated approach, crawling dataset metadata from the web and aggregating metadata from various sources, to provide a comprehensive dataset search engine. However, this search engine primarily reflects the perspectives of data publishers, potentially omitting the real-world application of datasets. DataCite~\cite{neumann2014datacite} assigns Digital Object Identifiers (DOIs) to datasets, improving their citability and accessibility. PapersWithCode (PwC)~\cite{martinez2021research} bridges academic publications with their associated code and datasets, fostering reproducibility. These platforms enhance transparency in the research ecosystem by systematically linking papers to underlying data. However, their reliance on manual annotations often results in incomplete dataset usage labels, limiting their comprehensiveness. Our work addresses the limitations of manual annotation by developing a self-evolving system that automatically extracts paper-dataset relationships from newly published papers.

\subsection{Information Extraction}
\label{sec:bg_dataset_extraction}
Information Extraction (IE) is a fundamental task of identifying and converting specific details, like named entities and their relationships, from unstructured or semi-structured text into a structured format~\cite{10.1145/234173.234209, li-etal-2023-extracting}. Traditionally, IE depends on supervised learning methods, which require a large amount of labeled data. With more weak supervision methods proposed~\cite{lison-etal-2020-named, 10.1145/3534678.3539247}, the need for annotation is alleviated. Recently, LLMs like GPTs~\cite{achiam2023gpt} have subverted the previous modeling methods of natural language processing tasks. For the IE problem, researchers have begun to explore zero-shot or few-shot learning techniques based on LLMs as a uniform tool~\cite{wei2021finetuned, ma-etal-2022-decomposed, li-etal-2023-type-aware, NEURIPS2024_5635925c}. Our work advances this paradigm by integrating LLMs to automate dataset information extraction, enhancing the scalability of detecting dataset usage in scholarly literature.

\subsection{Entity Resolution}

Entity Resolution (ER) is to identify multiple data representations of the same real-world entity and map them to a unified entity. The early ER methods are mainly based on distance-based methods, like edit distance method~\cite{monge1996field} and TF-IDF method~\cite{cohen2000data}. To overcome the limitations of unsupervised distance-based methods, researchers have proposed supervised learning methods. Ravikumar et al.~\cite{ravikumar2012hierarchical} define ER as a classification problem and use SVM to solve it. However, these methods are heavily based on labeled data. Recently, researchers have proposed unsupervised learning methods for ER. Lacoste-Julien et al.~\cite{lacoste2013sigma} propose a greedy matching method \textit{SiGMa} and Wu et al.~\cite{wu2020zeroer} propose \textit{ZeroER}, which uses a Gaussian Mixture Model to learn the similarity distributions of matches and non-matches to solve ER.
However, supervised learning methods require a large amount of labeled data, and unsupervised learning methods heavily rely on blocking methods, which makes them difficult to transfer to our dataset entity resolution. We propose a rule-based graph inference method leveraging strong indicator fields as relational constraints. Our algorithm performs iterative graph completion through deterministic pattern matching and transitive inference, achieving accurate entity resolution without training data or predefined blocking schemes.

\section{Problem Formulation}
\label{sec:problem_formulation}

We aim to construct a paper-dataset network that captures the usage of datasets in academic papers. Formally, the paper-dataset network can be defined as a bipartite graph \(G = (P, E, R)\), where \(P\) is the set of papers, \(E\) is the set of dataset entities, and \(R\) is the relationships between papers and datasets. Each edge \(r_{i, j} \in R\) connects a paper \(p_i \in P\) to a dataset entity \(e_j \in E\), indicating that the paper \(p_i\) uses the dataset entity \(e_j\).
Specifically, two main issues need to be addressed to construct the paper-dataset network:
\begin{itemize}[leftmargin=10pt]
	\item \textit{Dataset information extraction}: extract the dataset usage information from the texts of given papers;
	\item \textit{Dataset entity resolution}: align diverse dataset descriptions with their corresponding dataset entities, where a dataset entity represents a specific dataset within the dataset database.
\end{itemize}

\subsection{Dataset Information Extraction}

For each paper \( p \in P\), we have its text \(T(p)\). The information extraction is to apply the function \(F\) (realized via a prompt-based query to an LLM) to obtain:
\[
D(p) \coloneqq F(T(p)) = \{d_{p, 1}, d_{p, 2}, \ldots, d_{p, n(p)}\} \subseteq D
\]
where \(d_{p, i}\) is a JSON object representing the \(i\)-th dataset description in paper \(p\), and \(n(p)\) is the number of dataset descriptions in paper \(p\).
Here is an example of a JSON object for a dataset description:

\begin{lstlisting}[language=json,firstnumber=0]
{
    "dataset name": ...,
    "paper title": ...,
    "dataset summary": ...,
    "data type": ...,
    "task": ...,
    "url": ...,
    ... // other metadata related to the dataset
}
\end{lstlisting}

\subsection{Dataset Entity Resolution}

Given the dataset descriptions \(D\) extracted from papers and an initial dataset entity database \(E_{\text{init}}\) (derived from PwC), the objective of Entity Resolution (ER) is to find a mapping \(M : D \rightarrow E\), where \(E = E_{\text{init}} \cup E_{\text{new}}\). Each dataset description \(d \in D\) is mapped to an entity \(e \in E\) if they refer to the same real-world dataset. The set \(E_{\text{new}}\) contains new dataset entities not present in \(E_{\text{init}}\).

Formally, let \(\mathcal{C} = \{C_1, C_2, \ldots, C_m\}\) be a partition of \(D\) into equivalence classes under the relation \(d_i \sim d_j\) (indicating \(d_i\) and \(d_j\) refer to the same dataset). The mapping \(M\) is defined as:
\[
\forall C_k \in \mathcal{C}, \forall d \in C_k, M(d) = \begin{cases} 
e \in E_{init} & \text{if } \exists\ e \in E_{\text{init}} \text{ s.t. } e \sim C_k, \\
e_{\text{new}} \in E_{\text{new}} & \text{otherwise.}
\end{cases}
\]
This ensures each cluster \(C_k\) aligns with an existing entity in \(E_{\text{init}}\) when possible; otherwise, a new entity \(e_{\text{new}}\) is registered in \(E_{\text{new}}\) if the cluster indeed refers to a new real-world dataset. The resolution process constructs a paper-dataset network by connecting papers \(p \in P\) to their used dataset entities \(M(d) \in E\) for all descriptions \(d \in D(p)\).

\section{System Design}

In this section, we introduce \method, a novel LLM-driven system designed to automate the construction of a paper-dataset network. By leveraging LLMs to extract dataset information from academic papers and perform entity resolution, \method dynamically links papers to their corresponding datasets, forming a structured network.
As illustrated in Fig.~\ref{fig:system_arch}, the architecture of \method is built upon three pivotal modules:

\begin{itemize}[leftmargin=10pt]
	\item \textit{Paper Collection:} Aggregates papers from academic platforms to form the system’s foundational corpus.
	\item \textit{Dataset Information Extraction:} Identifies and extracts dataset-related text from academic papers, leveraging LLMs to generate semi-structured metadata (e.g., dataset names, data types, and associated tasks).
	\item \textit{Dataset Entity Resolution:} Resolves variant mentions of the same dataset by aligning them to a canonical entity, thereby constructing a paper-dataset bipartite graph.
\end{itemize}

\subsection{Paper Collection}

In the first phase, we collect basic information about academic papers. ArXiv~\cite{arxiv_org_submitters_2024}, one of the largest academic paper platforms, hosts a rich repository of preprints of research papers and is open on the web\footnote{https://www.kaggle.com/datasets/Cornell-University/arxiv}.
In the current implementation of \method, we collect papers from arXiv, focusing on Artificial Intelligence in Computer Science (cs.AI), and use the ar5iv tool \cite{stamerjohanns2010transforming} to obtain the text-format papers. 

We emphasize that \method operates independently of academic platforms, requiring only the text of papers for analysis. For example, by leveraging open-source PDF processing tools such as PyPDF\footnote{https://github.com/py-pdf/pypdf}, \method can build a personalized local paper-dataset network directly from a user’s collection of PDF documents.
Currently, we select arXiv as our primary source as it is fully open-access, and the majority of AI papers now publish preprints on this platform.

\begin{figure}[t]
	\centering
	\includegraphics[width=0.8\linewidth]{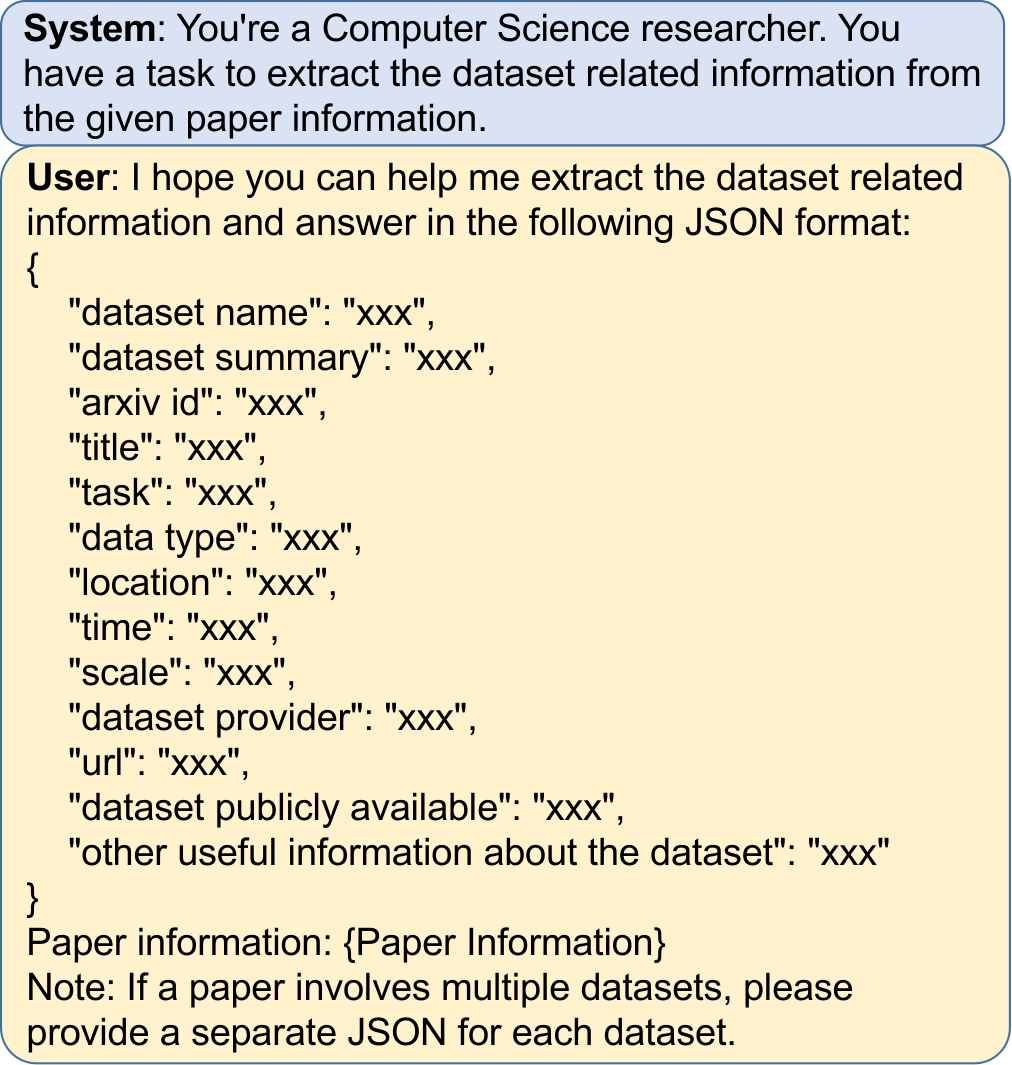}
	\caption{Dataset Information Extraction Prompt}
	\label{fig:dataset_extraction_prompt}
\end{figure}

\subsection{Dataset Information Extraction}
\label{sec:dataset_extraction}

The Dataset Information Extraction module identifies and extracts dataset-related metadata from academic papers collected in the preceding stage. For a paper \( p \), the module outputs a collection of dataset descriptions \( D(p) = \{d_{p, 1}, d_{p, 2}, \ldots, d_{p, n(p)}\} \), where each \( d_{p, i} \) represents a semi-structured JSON object encapsulating core dataset attributes.  

Recently, LLMs have shown great effectiveness and efficiency in analyzing text corpus \cite{min2023recent}. Based on LLMs, we can directly use chat-style natural interaction to extract useful dataset information from paper texts collected. With LLMs, there are three issues needed to be carefully considered: (1) \textit{prompt design}, (2) \textit{output quality control}, and (3) \textit{cost optimization}.

\subsubsection{Prompt Design} LLMs, e.g., ChatGPT, have recently showcased impressive performance in zero-shot or few-shot text information extraction tasks. 
To initiate the dataset information extraction process and generate responses in our desired structured format, we provide a specific prompt. The example of our prompt and corresponding demonstration is shown in Fig.~\ref{fig:dataset_extraction_prompt}. 

\textbf{Role}. Prior research has shown that specifying the role for the LLM would significantly improve the LLM's capability of solving the task \cite{zhang2023visar}.
Following the common practice, we set the role of the LLM as a computer science researcher, allowing it to better understand the task scenario. 

\textbf{Paper Information}.  The prompt features a `\textit{\{Paper Information\}}' field designed to incorporate relevant text from the paper pertaining to the dataset. Intuitively, this field could contain the entire paper text; however, in practice, this may result in prohibitively high costs when using LLM APIs, as computational expenses scale directly with input length. We explore this cost consideration in greater detail in Sec.~\ref{sub:prompt_cost}.

\textbf{Output Specification}. We also give specific task requirements and format standards for output. Previous research has summarized key considerations for researchers when finding datasets \cite{koesten2020everything}. We base our dataset information extraction on these key fields, such as the dataset name, data type, task, location, time, scale, and dataset providers. In addition to these key fields, we include the dataset summary, Uniform Resource Locator (URL), and other relevant information fields to offer a more comprehensive dataset description. To ensure the LLM produces semi-structured data, we instruct it to generate the output in JSON format. Considering that a paper may involve multiple datasets, we also add an annotation to remind the LLM to generate a JSON format description for each dataset.

\subsubsection{Output Quality Control} 

The ideal output would be standard JSON-formatted data for downstream processing.
However, our experiments reveal that even state-of-the-art LLMs (e.g., GPT-4o) occasionally generate outputs violating JSON syntax requirements. To mitigate this issue and ensure system reliability, we implement a dedicated format validation and correction step in the pipeline.
Specifically, we summarize three principal anomalies and institute corresponding rectifications via a post-processing script:

\begin{itemize}[leftmargin=10pt]
	\item \textit{Extraneous Expressions}: Entries not commencing with `\verb|{|', `\verb|}|', or `\verb|"|' are excised to eliminate non-pertinent phases.
	\item \textit{Malformed Escape Sequences}: We identify characters that need to be escaped in the output and add corresponding escape characters for them.
	\item \textit{Inconsistent Comma Usage}: We program to correct the problem of commas at the end of the line according to the syntax of JSON.
\end{itemize}

\subsubsection{Cost Optimization}
\label{sub:prompt_cost}
As we constrain the output to a JSON format with pre-defined fields, the cost of an LLM query is mostly related to the input length in \method.
In particular, the length of the paper text in the query, i.e., `\textit{\{Paper Information\}}', dominates the input length.
If we directly send the full paper text to LLM for processing, the cost would be relatively high especially when we want to scale up \method to deal with millions of papers.
To address this issue, we opt to input only the text of the paper sections that probably contain dataset descriptions. Academic papers usually describe the datasets used in the experimental sections, so we select sections like ``\textit{Experiment}'', ``\textit{Dataset description}'', ``\textit{Data}'', and other similar ones.
Considering the balance between the API call cost and the LLM's processing power, the length of the truncated input text is 1500 tokens (approximately 1125 words). Additionally, we include the title and abstract of the paper as supplementary input to provide a more comprehensive context of the datasets.

In our current implementation, the dataset information extraction module employs GPT-4o-mini\footnote{https://chat.openai.com/}, OpenAI's most advanced and cost-effective small-scale model. After cost optimization, the expense for \method to process 10,000 papers would be reduced to just \$6.3.
It is important to note that \method is not restricted to specific LLM services, and we have also evaluated other LLM services in our experiments. With the advancement of LLM techniques, we believe that it will soon be feasible to develop a fully local version of \method on a standard PC equipped with a mid-range graphics card. Exploring the deployment of such a locally deployable LLM model will be a focus of our future work.

\subsection{Dataset Entity Resolution}

The output of the dataset information extraction module is a set of dataset descriptions in JSON format, extracted from the paper texts. To construct the paper-dataset network, the next step is to extract dataset entities from these JSON-formatted descriptions. Specifically, there are two key challenges to address:

(1) \textbf{Existing Entity Matching}: When a paper uses a dataset that has already been referenced in other papers (i.e., an existing dataset entity in the database), the challenge is to correctly map the JSON-formatted description to the corresponding entity.

(2) \textbf{New Entity Discovery}: When a paper introduces a new dataset, the challenge is to identify it and register it as a new entity in the database.

\subsubsection{Existing Entity Matching}
\label{sub:existing_entity_match}

To initialize the dataset entity database, we currently utilize the dataset entities collected by the PwC platform. Through crowdsourcing, the PwC platform has accumulated a substantial number of dataset entities in its database, which include rich metadata such as dataset names and URLs. Additionally, PwC data is publicly accessible under the CC-BY-SA-4 license.\footnote{https://github.com/paperswithcode/sota-extractor} Our goal is to map the extracted dataset descriptions to their corresponding entities in the PwC database, thereby constructing a paper-dataset network.

In Sec.~\ref{sec:dataset_extraction}, we extract dataset-related information from paper texts, with certain fields—such as "dataset name" and "URL"—that can be used to identify the same dataset entity in the database. Our approach is based on the idea that if a dataset description shares the same name or URL as an existing dataset entity, we can conclude that the description refers to that entity with high confidence.

Following the idea, we propose a `\textbf{dataset identity attribute-based graph inference and completion}' algorithm to match dataset descriptions to existing entities.
First, we model the extracted dataset descriptions and database entities as nodes in a graph, referred to as \textit{description nodes (D-nodes)} and \textit{entity nodes (E-nodes)}, respectively. We then introduce \textit{identity-attribute nodes (I-nodes)} to represent unique identifiers such as dataset names and URLs. Notably, we create only one I-node for each unique dataset name or URL to avoid duplication. Next, we connect each I-node to its corresponding D-nodes and/or E-nodes. Then we introduce the graph \textit{inference} and \textit{completion} one by one.

\textbf{Graph Inference}: This graph structure enables us to infer relationships between D-nodes (dataset descriptions) and E-nodes (dataset entities).
For instance, if a D-node \(d\) is linked to an I-node and this same I-node is also connected to an E-node \(e\), we can infer that \(d\) corresponds to \(e\). This process effectively matches the dataset description to an existing dataset entity in the database through their shared identifier (e.g., same dataset name or URL).

Using the above process, we can match a D-node to its corresponding E-node if they share a common I-node. However, the original E-node in the database may initially connect to only a limited number of I-nodes, which restricts the coverage of this basic inference strategy. To address this limitation, we introduce a graph completion step to systematically enrich E-nodes’ connections to additional I-nodes, thereby improving inference coverage.  

\textbf{Graph Completion}:  
When a D-node \(d\) is matched to an E-node \(e\), all I-nodes connected to \(d\) are also linked to \(e\). This enriches \(e\)’s identity attributes by expanding its associated identifiers. Crucially, whenever a new I-node is connected to \(e\), we rerun the graph inference process for \(e\) to identify any additional D-nodes that can now be matched to \(e\) through the updated connections.  

\begin{algorithm}[t]
\footnotesize
	\caption{Graph Creation and Completion}
		\begin{algorithmic}[1]
		\State \textbf{Input:} A list of dataset descriptions $D = \{d_1, d_2, \ldots, d_n\}$, a list of entities $E = \{e_1, e_2, \ldots, e_m\}$
		\State \textbf{Output:} A graph $G = (V, \mathcal{E})$ with completions and corrections
		\State Identity attributes: $A = \{\text{dataset name}, \text{dataset url}\}$
            \State Initialize nodes: $V \gets D \cup E \cup \{I_{d, \alpha} \mid d \in D, \alpha \in A\} \cup \{I_{e, \alpha} \mid e \in E, \alpha \in A\}$

		\Comment{Graph Creation}
		\State $\mathcal{E} \gets \bigcup_{d\in D} \{(d \xrightarrow{\text{has\_}\alpha} I_{d, \alpha}) \mid \alpha \in A\}$

		\State $\mathcal{E} \gets \mathcal{E} \cup {\bigcup}_{e\in E} \{(I_{e, \alpha} \xrightarrow{\text{refers\_to}} e) \mid \alpha \in A\}$

		\While{iteration\_limit is not reached}
		\Comment{Graph Completion}

		\For{D-node $d \in D$}
		\For{attribute $\alpha \in A$}
		\If{$\exists\ \text{I-node}\ I_{d, \alpha} \xrightarrow{\text{refers\_to}} \text{E-node}\ e$}
			\State $\mathcal{E} \gets \mathcal{E} \cup \{(I_{d, A\setminus\{\alpha\}} \xrightarrow{\text{refers\_to}} e)\}$ 
		\EndIf
		\If{$|\{I_{d, \alpha} \xrightarrow{\text{refers\_to}} e\}| > 1$} \Comment {Refinement after Completion}
                \State Remove the I-node $I_{d, \alpha}$ and its edges from $V$ and $\mathcal{E}$
		\EndIf
		\EndFor
		\EndFor
		\EndWhile
    \State \Return $G$
	\end{algorithmic}
  \label{alg:graph_completion}
\end{algorithm}

Consider an E-node \(e_{\text{coco}}\) representing the MS COCO dataset~\cite{lin2014microsoft}, which initially has two I-nodes: the name “MS COCO” and the URL “https://cocodataset.org/”. During the inference step, we identify a D-node that shares the URL I-node but has an additional name I-node, “COCO 2014”. Through the graph completion step, we link the “COCO 2014” I-node to \(e_{\text{coco}}\). This enriched connection enables subsequent D-nodes associated with the “COCO 2014” I-node to be matched to \(e_{\text{coco}}\), thereby expanding the inference coverage.  

Considering the completion order, some I-nodes may not be connected to any E-node after the initial inference. To address this, we introduce a completion iteration to enrich the connections. In practice, we set an iteration limit to 3.

\textbf{Refinement after Completion}: While the graph completion strategy improves inference coverage, it risks introducing erroneous connections. A core principle is that I-nodes—representing identity attributes—should link to \textit{at most one E-node}. However, after completion, an I-node might connect to multiple E-nodes. This issue frequently arises with URL I-nodes. For instance, papers may cite generic data warehouse URLs like “www.kaggle.com” for used datasets, causing this I-node to link to multiple E-nodes for datasets hosted on Kaggle. Since such ambiguous I-nodes cannot reliably serve as unique identifiers, our current implementation of \method removes them from the graph to preserve integrity.

\begin{algorithm}[t]
\footnotesize
  \caption{Graph Inference for Entity Resolution}
  \begin{algorithmic}[1]
    \State \textbf{Input:} A list of dataset descriptions $D$, a list of dataset entities $E$, the completed graph $G = (V, \mathcal{E})$
    \State \textbf{Output:} A list of matched dataset descriptions and entities $M$
    \State $M \gets \{\}$
    \For{D-node $d \in D$}
	\For{attribute $\alpha \in $\{dataset name, dataset url\}}
	\If{$\exists\ \text{I-node}\ I_{d, \alpha} \xrightarrow{\text{refers\_to}} \text{E-node}\ e$}
	\State $M \gets M \cup \{(d, e)\}$
	\EndIf
	\EndFor
    \EndFor
    \State \Return $M$
  \end{algorithmic}
  \label{alg:entity_resolution}
\end{algorithm}

After graph completion and refinement, we can infer the final mappings between dataset descriptions (D-nodes) and their corresponding entities (E-nodes) in the database. The full process is formalized in Algorithm~\ref{alg:graph_completion} and \ref{alg:entity_resolution}.

\subsubsection{New Entity Discovery}
\label{sec:extended_database}

Another key strength of \method lies in its ability to discover novel dataset entities from academic literature. For example, our analysis reveals that nearly \textit{50\% of datasets} extracted by \method from arXiv papers are absent from PwC’s database, highlighting these datasets' novelty and suggesting they represent emerging resources useful for academic research.

After the graph inference and completion (Sec.~\ref{sub:existing_entity_match}), some D-nodes may remain unmatched to any E-nodes. These unmatched D-nodes could represent novel dataset entities introduced by the corresponding papers. However, automatically creating a new E-node for every unmatched D-node risks introducing noise, as dataset descriptions extracted by LLMs may contain inaccuracies. To address this, \method enforces two criteria to determine whether an unmatched D-node needs the creation of a new E-node.  

1. \textbf{Identity Information Completeness}. Currently, \method only considers creating E-nodes for unmatched D-nodes with complete identity attributes, i.e., containing both a dataset name and a URL. Notably, after graph refinement (Sec.~\ref{sub:existing_entity_match}), all URL I-nodes associated with generic data warehouse links (e.g., “www.kaggle.com”) are removed. Therefore, if an unmatched D-node retains a URL I-node, it is likely a \textit{specific}, \textit{non-generic URL}, increasing confidence that the D-node represents a genuinely new dataset. 

2. \textbf{Multiple Paper Mentions}.  \method prioritizes creating new E-nodes when \textit{multiple unmatched D-nodes} share identical I-nodes (e.g., the same dataset name or URL). This increases confidence that the dataset is genuine and significant, as it is independently mentioned across multiple papers. For such cases, \method consolidates all D-nodes sharing same I-nodes into a \textit{single} E-node, representing one unified novel dataset entity.
In the implementation, We can define a threshold \(\lambda\) to govern the creation of new E-nodes: a candidate dataset must be mentioned in at least \(\lambda\) papers. 

Additionally, we plan to incorporate \textit{user feedback} to improve the accuracy and efficiency of dataset discovery. For example, even if a dataset lacks mentions from \(\lambda\) papers, we still create an E-node but flag it with an \textit{uncertainty indicator}. When presenting such datasets, \method could ask users to verify dataset accuracy (e.g., “Is this extracted dataset correct?”). User feedback, while valuable, is not always reliable. Accurately extracting trustworthy insights from such feedback remains a significant challenge—a problem widely recognized in literature as truth discovery. We defer addressing this challenge to future research.

\section{Experiment}

We evaluate \method to ascertain its effectiveness in constructing the paper-dataset network following three questions:

\begin{itemize}[leftmargin=10pt, label={}]
\item \textit{RQ1: Can \method efficiently and accurately extract dataset information?}
\item \textit{RQ2: Can \method effectively resolve dataset descriptions entities?}
\item \textit{RQ3: Can \method discover new datasets?}

\end{itemize}

\subsection{Performance of Dataset Information Extraction (RQ1)}

\subsubsection{Experimental Setup}

To compare \method, we implement three comparative approaches:

(1) \textbf{en\_core\_web\_trf}: employing a named entity recognition model en\_core\_web\_trf \footnote{\url{https://spacy.io/models/en\#en_core_web_trf}} to detect dataset entities in papers\cite{prasad2019dataset}.\\
en\_core\_web\_trf is a powerful pre-trained transformer-based model that can recognize and label a variety of entities in text, including dataset names \cite{liu2019roberta}.

(2) \textbf{Regular Expression}: using regular expressions to identify and match dataset names and their common variants in paper text based on a predefined list of dataset names (e.g., hyphenation variations like "Mini-ImageNet" and "MiniImageNet")\cite{pan2023dmdd}.

(3) \textbf{PapersWithCode (PwC)}: directly using the datasets identified by \textit{PwC} for test papers. The dataset usage information on \textit{PwC} is derived partly from annotations by community members and partly from the rule-based automated extraction script.\footnote{\url{https://github.com/paperswithcode/sota-extractor}} 

For implementing LLM APIs in \method, we choose GPT-3.5-turbo, GPT-4o-mini (default), Qwen2.5-7b-instruct \cite{qwen2.5}, and DeepSeek-V3 \cite{liu2024deepseek} for comparison. To compare with our cost optimization strategy (Sec.~\ref{sub:prompt_cost}), we also implement a variant of inputting the full paper text to LLMs.

To construct the test set, we manually annotate datasets used in research papers to establish a ground truth for evaluation. Specifically, we annotate dataset usage in 119 papers from top-tier conferences, including KDD and NeurIPS. The statistics of the annotated papers are detailed in Table~\ref{tab:dataset_annotation}.
To ensure a fair comparison with PwC, our selected test papers all have dataset annotations on PwC.

\begin{table}[t]
\footnotesize
	\centering
	\caption{Dataset Usage Statistics in Annotated Papers}
	\begin{tabular}{lccc}
		\toprule & \textbf{NeurIPS} & \textbf{KDD} & \textbf{Total} \\
		\midrule
		\textbf{\# Papers}           & 50               & 69           & 119           \\
		\textbf{\# Total Datasets Used}        & 110              & 186          & 296            \\
		\textbf{\# Unique Datasets Used}       & 61               & 97          & 143            \\
		\textbf{Avg. \# Datasets Per Paper} & 2.20             & 2.70         & 2.49           \\
		\bottomrule
	\end{tabular}
	\label{tab:dataset_annotation}
\end{table}

\subsubsection{Results}

\begin{figure}[t]
	\centering
	\includegraphics[width=0.49\textwidth, height=0.18\textheight]{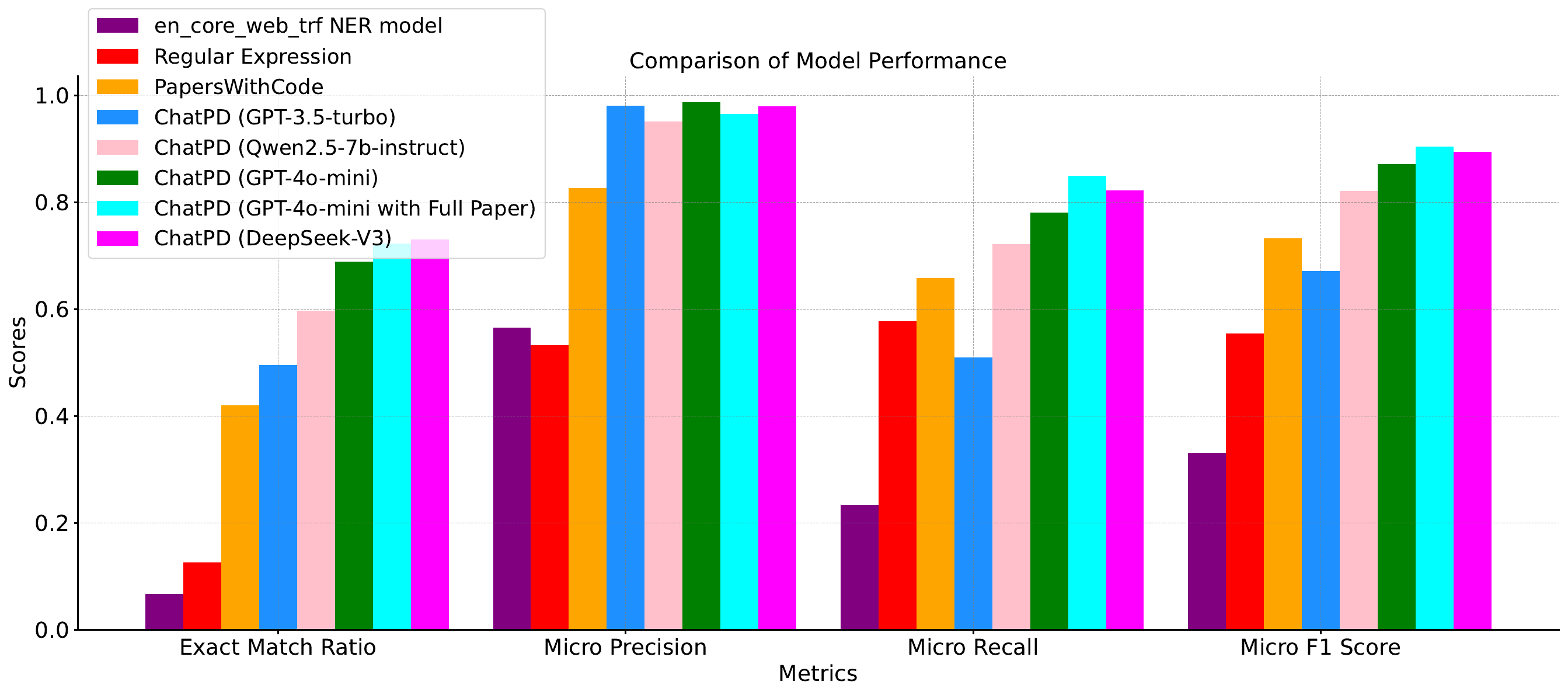}
	\caption{Performance of Dataset Information Extraction}
	\label{fig:dataset_extraction_performance}
\end{figure}

We evaluate the performance of dataset information extraction by calculating various metrics, including \textit{Exact Match Ratio}, \textit{Micro Average Precision}, \textit{Micro Average Recall} and \textit{Micro Average F1 score}. 
The comparison results are shown in Fig.~\ref{fig:dataset_extraction_performance}.

Our results indicate that \textit{Regular Expression}
and \textit{en\_core\_web\_trf} struggle to effectively capture dataset information. \method with GPT-3.5-turbo achieves competitive performance compared with \textit{PwC}. With more advanced LLMs such as GPT-4o-mini and DeepSeek-V3, \method outperforms \textit{PwC} significantly across all metrics.
Our method remains robust even with lightweight, locally deployable models such as Qwen2.5-7b-instruct.
By analyzing the data, we observe that the unsatisfactory performance of \textit{PwC} can be attributed to its rule-based extraction technique for identifying datasets from texts. This method frequently results in erroneous matches, e.g., wrongly identifying datasets that are merely referenced in the text but not actually used in the study. 

To evaluate the effectiveness of our cost optimization strategy, we conduct a comparison between the full-text input and our optimized 1500-token input using GPT-4o-mini. The results demonstrate that the 1500-token input achieves performance close to the full-text input, and even outperforms it in certain metrics like Precision. Note that processing the full text would require approximately 7 times more tokens compared to our optimized method, significantly increasing costs. Given that \method is designed to handle a continuous and large volume of papers, we believe that limiting the input to 1500 tokens strikes an effective balance between cost efficiency and performance.

Overall, our experimental results show that \method with current LLMs are highly effective in extracting datasets from papers, surpassing state-of-the-art solutions like \textit{PwC}, highlighting the feasibility of using large language models for this task.

\subsection{Performance of Dataset Description Entity Resolution (RQ2)}
\label{sec:entity_resolution_performance}

\subsubsection{Experimental Setup}

In this experiment, we aim to match dataset descriptions to existing dataset entities. Specifically, we utilize the dataset entities already stored in PwC as the reference existing entities. To establish ground truths, we manually annotate dataset descriptions extracted from papers published in top-tier conferences, such as KDD and NeurIPS, by linking them to their corresponding entities in the database.

We sample 1,000 dataset descriptions randomly and link them manually to the corresponding entities. We find that only 474 dataset descriptions, only half of the samples, can be linked to the dataset entities in the PwC database.
The primary reason for the unlinked descriptions is the absence of corresponding entities in the PwC database. Additionally, some descriptions were too vague, such as `weather dataset', to determine their corresponding entities.

We compare our \textit{Graph Completion \& Inference} algorithm with the \textit{Name Matching} method (connecting descriptions to entities with the same dataset name) and the \textit{Graph Inference} algorithm (connecting dataset descriptions with the same dataset name, alias, or URL). Besides, we compare two popular entity resolution algorithms, \textit{SiGMa}~\cite{lacoste2013sigma} and \textit{ZeroER}~\cite{wu2020zeroer}.

\subsubsection{Results}

We choose precision, recall, and F1 score as the evaluation metrics. Our results are shown in Table~\ref{tab:entity_resolution}. 

\begin{table}[t]
\footnotesize
	\centering
	\caption{Evaluation of Entity Resolution Methods}
	\label{tab:entity_resolution}
	\setlength{\tabcolsep}{1.0mm}{
		\begin{tabular}{lccc}
			\hline
			\textbf{Method}     & \textbf{Precision} & \textbf{Recall} & \textbf{F1 Score}\\
			\hline
			Name Matching & \textbf{1.0000} & 0.5105 & 0.5680 \\
			SiGMa \cite{lacoste2013sigma} & 0.7319 & 0.6312 & 0.6778 \\
			ZeroER \cite{wu2020zeroer} & 0.9984 & 0.5844 & 0.6300 \\
			Graph Inference & 0.9917 & 0.6477 & 0.7007 \\
			Graph Completion \& Inference & 0.9826 & \textbf{0.8727} & \textbf{0.8829} \\
			\hline
		\end{tabular}
	}
\end{table}

\textit{Name Matching} achieves the highest precision, but cannot find the same dataset with different names, leading to the lowest recall. As a result, its F1 score is also the worst. 
\textit{Graph Inference} utilizes the aliases and URLs provided by PwC, achieving a higher recall and F1 score than the state-of-the-art methods \textit{SiGMa} and \textit{ZeroER}.

Our \textit{Graph Completion \& Inference} algorithm considers the transitive relationship between dataset descriptions, which can further increase the recall. It achieves the best F1 score $0.8829$, verifying its effectiveness in constructing the paper-dataset network.

\subsection{New Dataset Entity Discovery (RQ3)}
\label{sub:new_dataset_results}
By applying the new dataset entity discovery strategy (Section~\ref{sec:extended_database}), \method can detect novel dataset entities referenced in academic papers.
We list the top 10 most frequently used new dataset entities discovered by \method that were not included in PwC's dataset database as of November 16, 2024. We compare the coverage of these dataset entities in PwC's database on November 16, 2024, and January 16, 2025. The results are shown in Table~\ref{tab:new_dataset_entity_discovery}. 

Only three out of the ten popular new datasets were added to PwC as of January 16, 2025. Notably, the most widely used dataset, \textit{UltraFeedback} \cite{cui2023ultrafeedback}, which has been used in over 40 papers, is still not included in PwC. This highlights that \method is significantly more efficient at discovering new dataset entities compared to PwC.

\begin{table}
	\caption{New Dataset Entities Discovered by \method}
	\label{tab:new_dataset_entity_discovery}
	\centering
	\resizebox{0.5\textwidth}{!}{
	\begin{tabular}{lcccc}
		\toprule
		\textbf{Dataset Entity} & \textbf{Usage Count} & \textbf{ChatPD} & \textbf{PwC (2024.11.16)} & \textbf{PwC (2025.01.16)} \\
		\midrule
		UltraFeedback\cite{cui2023ultrafeedback} & 43 & \checkmark & - & - \\
		Diginetica\cite{Diginetica} & 22 & \checkmark & - & - \\
		BabyAI\cite{chevalier2018babyai} & 19 & \checkmark & - & - \\
		HELOC\cite{heloc2018} & 16 & \checkmark & - & \checkmark \\
		RedPajama\cite{together2023redpajama} & 16 & \checkmark & - & - \\
		Camelyon17\cite{bandi2018detection} & 15 & \checkmark & - & - %(Camelyon16\cite{litjens_1399_2018} included) 
        \\
		VirtualHome\cite{puig2018virtualhome} & 15 & \checkmark & - & - \\
		California Housing\cite{ca_housing} & 14 & \checkmark & - & \checkmark \\
		Yoochoose\cite{ben2015recsys} & 14 & \checkmark & - & - \\
		Titanic\cite{titanic} & 13 & \checkmark & - & \checkmark \\
		\bottomrule
	\end{tabular}
	}
\end{table}

\section{Deployment}
\label{sec:deploy}

\method has been deployed to update the paper-dataset network weekly. Users can access \textbf{\url{https://chatpd-web.github.io/chatpd-web}} to search for datasets used in papers by specifying the arXiv ID or dataset name, data type, task, etc. 
We present the basic dataset services provided by \method after deployment in Appendix~\ref{sec:dataset_discovery_service}.

\subsection{Offline Results}
\label{sec:network_statistics}

Before deployment, we conduct offline evaluations to ensure the effectiveness and efficiency of \method.
We randomly sample 35,310 papers in the cs.AI category on arXiv and extract dataset information from them by \method.
We compare the data extracted by \method with that from the platform \textit{PwC} to analyze the network's size and coverage.

Table \ref{tab:network_coverage} provides a summary of the network size and coverage metrics for \textit{PwC} and \method.

\begin{table}[t]
\footnotesize
	\centering
	\caption{Network Size and Coverage Statistics}
	\begin{tabular}{lcc}
		\toprule
		\textbf{Metric} & \textbf{PwC} & \textbf{ChatPD} \\
		\midrule
		\textbf{\# Papers Extracted} & 14,353 & 35,310 \\
		\textbf{\# Dataset Descriptions Extracted} & 32,146 & 76,056 \\
		\textbf{\# PwC Dataset Entities Covered} & 3,556 & 3,144 \\
		\textbf{\# Descriptions Matched to PwC Entities} & 32,146 & 35,085 \\
		\textbf{\# New Dataset Entities Discovered} & - & 444 \\
		\textbf{\# Descriptions Matched to New Entities} & - & 1,217 \\
		\textbf{Avg. Cost Per Paper Extracted (USD)} & - & 0.00063 \\
		\bottomrule
	\end{tabular}
	\label{tab:network_coverage}
\end{table}

The data indicates that \method has significantly expanded the scope of the paper-dataset network compared to \textit{PwC}. 
\method has extracted dataset usage information from more than double the number of papers and dataset descriptions compared to \textit{PwC}.

Besides the existing PwC entities, \method also find 444 new dataset entities not included in PwC. Specifically, we infer a new dataset entity if it has a useful URL and is referenced by at least 3 papers (Sec.~\ref{sec:extended_database}). Additionally, its cost-efficiency is notable, with an average cost per paper extracted being significantly low at \$0.00063 using GPT-4o-mini.

Through offline evaluation, we demonstrate that \method constructs a larger and more comprehensive paper-dataset network with impressive cost efficiency.

\subsection{Post-Deployment Results}

We evaluate the performance of the deployed \method by analyzing the paper-dataset network constructed from cs.AI papers on arXiv in 2024. Our results are summarized in Table~\ref{tab:performance_evaluation}.

\begin{table}[t]
\footnotesize
    \centering
	\caption{Performance Evaluation of \method in the cs.AI Category on arXiv (2024)}
    \resizebox{0.5\textwidth}{!}{
        \begin{tabular}{lcc}
            \toprule
            \textbf{Metric} & \textbf{Value} \\
            \midrule
            \textbf{\# Papers in arXiv cs.AI (2024)} & 32,959 \\
            \textbf{\# Papers with Accessible Text Information via ar5iv} & 28,901 \\
            \textbf{\# Papers with Dataset Information Successfully Extracted} & 24,719 \\
			% 论文的处理成功率
			\textbf{Success Rate of Paper Processing} & 85.5\% \\
            \textbf{\# Dataset Descriptions Extracted} & 59,664 \\
			\textbf{Avg. \# Dataset Usage Records Per Paper} & 2.41 \\
            \textbf{\# Descriptions Matched to PwC Entities} & 27,428 \\
            \bottomrule
        \end{tabular}
    }
    \label{tab:performance_evaluation}
\end{table}

Our results show that approximately 87.8\% of papers have accessible text information via ar5iv. \method successfully extracts dataset information from 85.5\% of these papers, with an average of 2.41 dataset usage records per paper.
Among the extracted dataset usage records, less than half of the dataset descriptions can be mapped to PwC's dataset entities. Our offline experiments in Section~\ref{sec:entity_resolution_performance} demonstrate the effectiveness of our entity resolution algorithm for mapping dataset descriptions to PwC's dataset entities. Hence, this low matching ratio indicates that PwC's database is still incomplete, i.e., there is still significant room for improvement in the coverage of PwC's dataset database.

\begin{figure}[t]
	\centering
	\includegraphics[width=0.45\textwidth]{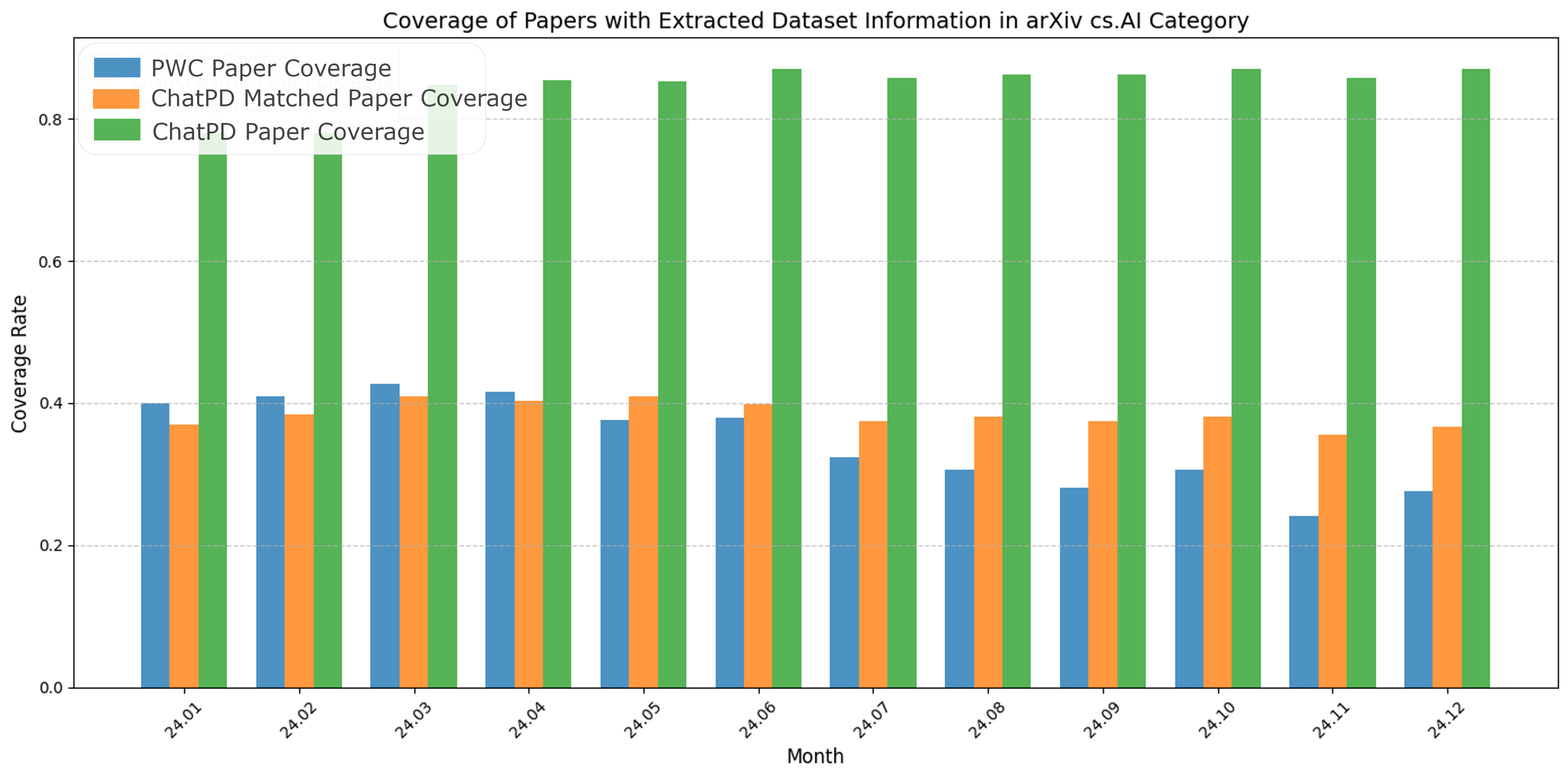}
	\caption{Coverage of Papers with Extracted Dataset Information in arXiv cs.AI Category}
	\label{fig:chatpd_pwc_month_cmp}
\end{figure}

We also evaluate the real-time performance of the deployed \method and compare it with PwC's results. We calculate the coverage of papers with extracted dataset information in the PwC database and the coverage of dataset usage records extracted by \method by month. The results are shown in Fig.~\ref{fig:chatpd_pwc_month_cmp}.
As not all the extracted dataset descriptions can find matching entities in the PwC database, we record both `the coverage of papers with matched PwC entities (\method Matched Paper Coverage)' and `the coverage of papers with extracted dataset information (\method Paper Coverage)'.
Our data is up to January 12, 2025.

We observe that PwC's paper coverage is higher than \method's matched paper coverage at the beginning of 2024. However, after May, \method's coverage surpasses PwC's. PwC's coverage is relatively low for newly published papers due to its partial reliance on community annotations. In contrast, \method uses LLMs to automatically extract dataset information, enabling it to stably analyze dataset usage records in papers. Therefore, \method's coverage is significantly higher than PwC's in the later months. In 2024, PwC's paper coverage is 34.5\%, \method's paper coverage that can be mapped to PwC dataset entities is 38.4\%, and the paper coverage with extracted dataset information is 85.5\%. This demonstrates that \method can stably and efficiently extract dataset information.

\section{Conclusion}

In this paper, we introduce and deploy a novel Large Language Model (LLM)-driven system, \method, for constructing a comprehensive paper-dataset network. \method automates the extraction of dataset information from academic papers, enabling the construction of a structured network that captures the intricate relationships between papers and datasets. Through our entity resolution algorithm, we effectively map diverse dataset descriptions to their corresponding real-world dataset entities.
We evaluate \method's performance in dataset information extraction, dataset description entity resolution, and network construction through offline experiments. We deploy \method on papers in the cs.AI category on arXiv and evaluate its deployment performance. We also demonstrate \method's dataset discovery services, including table-based and graph-based queries (Appendix \ref{sec:dataset_discovery_service}).

However, we must acknowledge that due to the current limitations of LLMs, there may be some errors in the fully automated construction of the paper-dataset network. We believe that our future system can collaborate with platforms like PwC to transition from entirely manual annotation to manual calibration based on the results obtained from \method. This can significantly reduce the workload of manual annotation and yield a more accurate paper-dataset network.

As we continue to refine and expand our system, we are optimistic about its potential to transform the way researchers interact with datasets, making the landscape of academic research more interconnected and accessible than ever before. 

\section*{Acknowledgements}

This work was supported by the National Natural Science Foundation of China (NSFC) under Grant No.~U23A20468.

\bibliographystyle{ACM-Reference-Format}
% \balance
\bibliography{references}

\appendix

\section{Appendix}

\subsection{Dataset Discovery Service}
\label{sec:dataset_discovery_service}

Building upon the constructed paper-dataset network, our system is poised to offer a suite of services designed to enhance the research community's ability to discover and utilize datasets effectively. These services are not only aimed at simplifying the dataset search process but also at providing insights into dataset relevance, usage trends, and their applicability to various research tasks. Here we divide the services supported by the current network into two categories: table-based query and graph-based query.

\subsubsection{Table-Based Query}

The table-based query is a traditional way in the database to query the dataset information. Users can search for datasets by specifying criteria such as the dataset name, associated tasks, data types, or the research domains they are interested in. The search results are augmented with information on how and where the datasets have been used in literature, offering researchers valuable context.

\textit{Demo: Which datasets occurred in New York?}
\par
Urban planners, for example, may want to know which datasets are used in New York. They can use something like Structured Query Language (SQL) query to get the answer.

\begin{lstlisting}[language=SQL, firstnumber=0, basicstyle=\footnotesize\ttfamily]
SELECT * FROM dataset WHERE location LIKE '%New York%'
\end{lstlisting}

Here is the result of the query:

\begin{lstlisting}[language=json,firstnumber=0,label=lst:dataset_newyork, basicstyle=\fontsize{5pt}{6pt}\selectfont]
{
"arxiv id": "2108.04462",
"title": "Deep Reinforcement Learning for Demand Driven Services in Logistics and Transportation Systems: A Survey",
"dataset name": "New York City TLC Dataset",
"dataset summary": "Contains travel records for various services, including Yellow taxis, Green taxis, and FHV (For-Hire Vehicle) from 2009 to 2020.",
"task": "Dispatching",
"data type": "Travel records",
"location": "New York City, USA",
"time": "2009-2020",
"scale": "Large",
"dataset provider": "New York City TLC",
"dataset url": "https://www.nyc.gov/site/tlc/about/tlc-trip-record-data.page",
"dataset publicly available": "Yes",
"other useful information about this dataset": "This dataset is commonly used for various studies in transportation and urban mobility."
},
...
\end{lstlisting}

\subsubsection{Graph-Based Query}

\begin{figure}[t]
	\centering
	\includegraphics[width=0.44\textwidth, height=0.23\textheight]{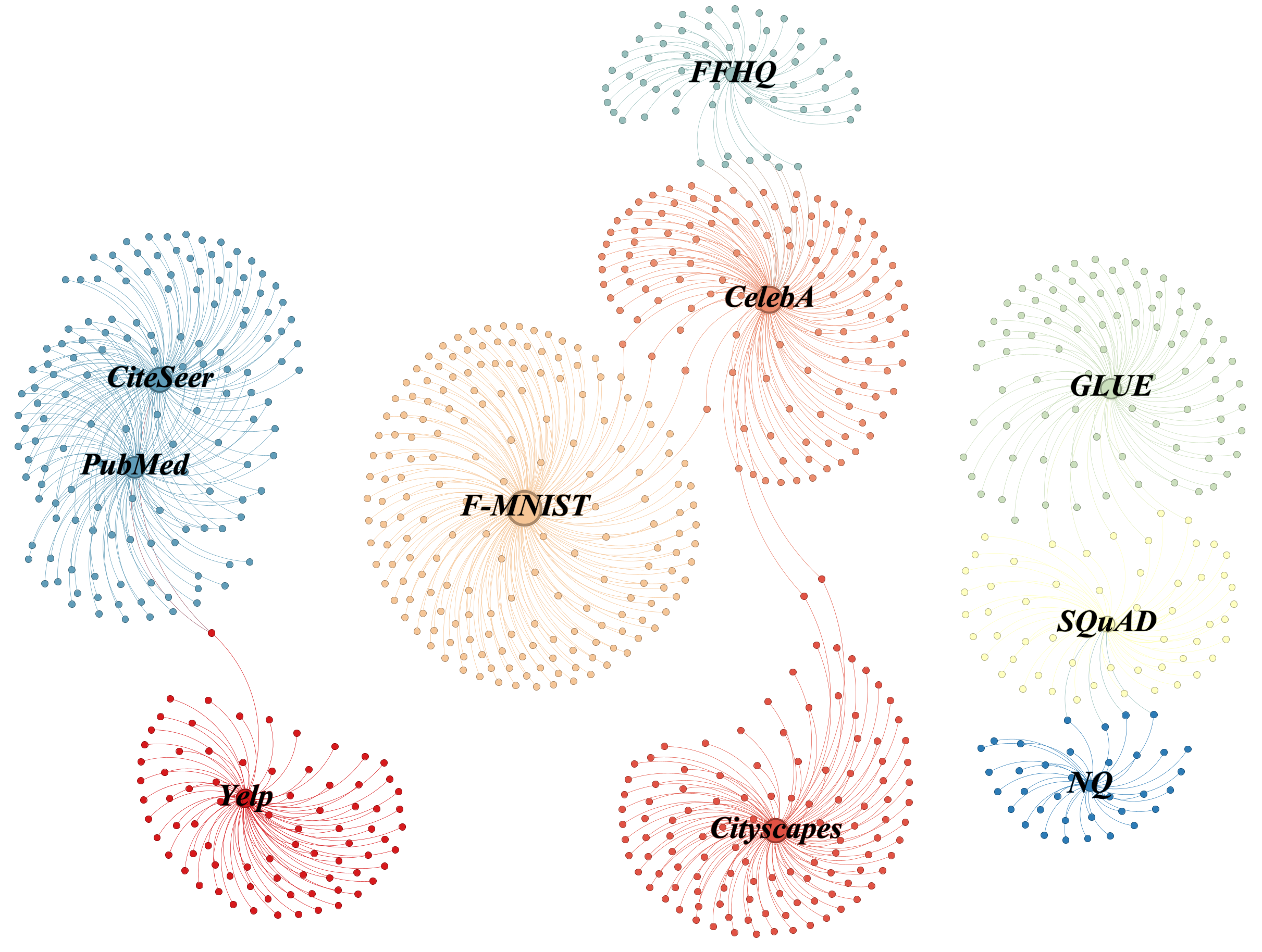}
	\caption{Visualization of Paper-Dataset Network (F-MNIST means Fashion-MNIST, NQ means Natural Questions).}
	\label{fig:network_visualization}
\end{figure}

Our paper-dataset network is a kind of data in graph form, which has rich information. We use the Gephi\cite{bastian2009gephi}, a popular network visualization tool, to visualize the constructed paper-dataset network. The network is a bipartite graph, where each node represents a paper or a dataset, and each edge represents the usage of a dataset in a paper. We sampled Cityscapes\cite{cordts2016cityscapes}, CelebA\cite{liu2015deep}, FFHQ\cite{karras2019style}, and Fashion-MNIST\cite{xiao2017fashion} for Computer Vision, GLUE\cite{wang2018glue}, SQuAD\cite{rajpurkar2016squad}, and Natural Questions\cite{kwiatkowski2019natural} for Natural Language Processing, and Yelp\cite{lin2017structured}, PubMed\cite{sen2008collective}, Cite\-Seer\cite{giles1998citeseer} for graph-based analyses.
As shown in Fig.~\ref{fig:network_visualization}, we can notice that different domain datasets are naturally clustered together.
This is also in line with common sense, that is, papers in different research directions will use multiple datasets in the field to conduct experiments, verifying the generalization of the proposed methods.

Inspired by this, we can use graph algorithms to find similar datasets to a given dataset.

One effective way to find similar datasets with the paper-dataset network is by employing a Random Walk with Restart (RWR) algorithm\cite{tong2008random}. The RWR algorithm is a graph-based algorithm that simulates the process of a random walker traversing the graph. The walker starts at a given node and moves to a neighboring node with a certain probability. The walker can also restart at the original node with a certain probability. The RWR algorithm is widely used in graph-based recommendation systems and is effective in finding similar nodes in a graph.

\textit{Demo: Which datasets are similar to dataset SQuAD?}

The Stanford Question Answering Dataset (SQuAD) is a dataset of question-answer pairs, which is widely used in the field of natural language processing. We can use the RWR algorithm to find datasets similar to the SQuAD dataset. The top 5 similar datasets are shown in Table~\ref{tab:similar_datasets}.

\begin{table}[t]
	\centering
	\caption{Top 5 Datasets Similar to the SQuAD}
	\label{tab:similar_datasets}
	\setlength{\tabcolsep}{4.5mm}{
		\begin{tabular}{lc}
			\hline
			\textbf{Dataset} & \textbf{Similarity Score} \\
			\hline
			GLUE             & 0.0198                    \\
			Natural Questions         & 0.0110                    \\
			NewsQA             & 0.0098                    \\
			TribiaQA         & 0.0097                    \\
			MS MARCO         & 0.0064                   \\
			\hline
		\end{tabular}
	}
\end{table}

Similar to the SQuAD dataset, the GLUE~\cite{wang2018glue}, Natural Questions~\cite{kwiatkowski2019natural}, NewsQA~\cite{trischler2016newsqa}, MS TriviaQA~\cite{joshi2017triviaqa}, and MS MARCO~\cite{nguyen2016ms} datasets are widely used for training and evaluating machine reading comprehension models. Researchers often employ the SQuAD dataset in conjunction with these datasets to verify whether their models' understanding and reasoning capabilities generalize across diverse benchmarks.

\subsection{Agentic Framework for Dataset Information Extraction}

\method currently selects sections of the paper that are relevant to dataset information extraction based on rules. However, this approach may overlook datasets that are used in sections other than those explicitly designated for dataset information extraction, such as the Related Work section.

To address this issue, we propose an Agentic Framework for dataset information extraction, inspired by the agent module in the HLM-Cite \cite{NEURIPS2024_5635925c} system, to better locate and extract dataset-related content scattered across different sections of the paper.
The framework is designed as follows:

\begin{itemize}
    \item Summarizer: Summarizes the main content of each section of the paper.
    \item Selector: Selects sections that are likely to contain dataset-related information based on their summaries.
    \item Extractor: Analyzes the selected sections to extract structured dataset details such as names, tasks, and data types.
\end{itemize}

We evaluate the performance of the Agentic Framework on the dataset information extraction task and supplement the cost comparison in Table~\ref{tab:extraction_comparison}. We use GPT-4o-mini as the LLM backend in \method and use the different input methods for comparison.

\begin{table}[t]
    \centering
    \caption{Comparison of Input Strategies for Dataset Information Extraction}
    \label{tab:extraction_comparison}
    % \small
    \begin{tabular}{lccccc}
        \hline
        \textbf{Input Strategy} & \textbf{EMR} & \textbf{P} & \textbf{R} & \textbf{F1} & \textbf{Cost} \\
        \hline
        PapersWithCode & 0.420 & 0.826 & 0.659 & 0.733 & - \\
        ChatPD (1.5k tokens) & 0.689 & 0.987 & 0.780 & 0.872 & 0.00063 \\
        ChatPD (Full Paper) & 0.723 & 0.966 & 0.850 & 0.904 & 0.00447 \\
        ChatPD (Agentic) & 0.773 & 0.896 & 0.946 & 0.920 & 0.00739 \\
        \hline
    \end{tabular}
    {\footnotesize
    \begin{flushleft}
    \textbf{Note:} EMR = Exact Match Ratio, P = Micro Precision, R = Micro Recall, F1 = Micro F1 Score. Cost is measured in USD per paper.
    \end{flushleft}
    }
\end{table}

We notice that the Agentic Framework performs better than the previous method in all metrics except for Micro Precision, indicating that the Agentic Framework can better locate dataset-related content and extract more comprehensive dataset usage information. By analyzing the bad cases, we find that the Agentic Framework sometimes mistakenly identifies datasets that were mentioned (but not actually used) in the Related Work section as the dataset used by the paper, which leads to a decrease in Micro Precision.

The Agentic Framework demonstrates strong overall performance and promising potential, but currently costs about 11.7 times more than the original method. In the future, we will explore a more efficient Agentic Framework to optimize the current \method.

\end{document}